\documentstyle[pasjskel,graphicx,natbib,twocolumn]{article}

\def\inpress{in press}
\def\astroph#1{ (astro-ph/#1)}
\def\labelspace{}

\def\AcA{Acta Astron.}
\def\ASPConf#1#2{ASP Conf. Ser. #1, #2}
\def\PublisherASP{San Francisco: ASP}

\begin{document}

\title{On the Origin of Early Superhumps in WZ Sge-Type Stars}

\author{Taichi \textsc{Kato}}
\affil{Department of Astronomy, Kyoto University,
       Sakyo-ku, Kyoto 606-8502}
\email{tkato@kusastro.kyoto-u.ac.jp}


\begin{abstract}
   The origin of {\it early superhumps}, which are double-wave periodic
modulations seen only during the earliest stage of WZ Sge-type outburst,
has not been well understood.  Based on recent discovery of two-armed
arch-like patterns on Doppler tomograms in conjunction with
early superhumps, we propose a new interpretation on the origin of early
superhumps, following the new interpretation by \citet{sma01tidal} and
\citet{ogi01tidal} of the two-armed pattern seen in IP Peg.  If we consider
irradiation of the elevated surface of the accretion disk caused by
vertical tidal deformation, we can consistently explain the observed
features on Doppler tomograms and photometric waves at the same time.
We interpret that a combination of low mass-ratios ($q$) and low
mass-transfer rates, necessary to give rise to these tidal effects,
discriminate WZ Sge-type stars from other SU UMa-type dwarf novae.
Based on recent fluid calculations, such an effect would be observable
in higher $q$ systems.  We interpret that RZ Leo is an example of such
objects.
\end{abstract}

Key Words: accretion, accretion disks
          --- stars: novae, cataclysmic variables
          --- stars: dwarf novae
          --- stars: individual (WZ Sagittae, RZ Leonis)

\section{Introduction}
   Dwarf novae are a class of cataclysmic variables (CVs), which are
close binary systems consisting of a white dwarf and a red dwarf secondary
transferring matter via the Roche-lobe overflow.  WZ Sge-type dwarf
novae (cf. \cite{bai79wzsge}; \cite{dow81wzsge}; \cite{pat81wzsge};
\cite{dow90wxcet}; \cite{odo91wzsge}; \cite{kat01hvvir}) are a small
subgroup of dwarf novae characterized by the long ($\sim$ 10 yr) outburst
recurrence time and the large ($\sim$ 8 mag) outburst amplitude.
They are a subclass of SU UMa-type dwarf novae (cf. \cite{war95suuma}).

\begin{figure}
  \begin{center}
    \FigureFile(80mm,110mm){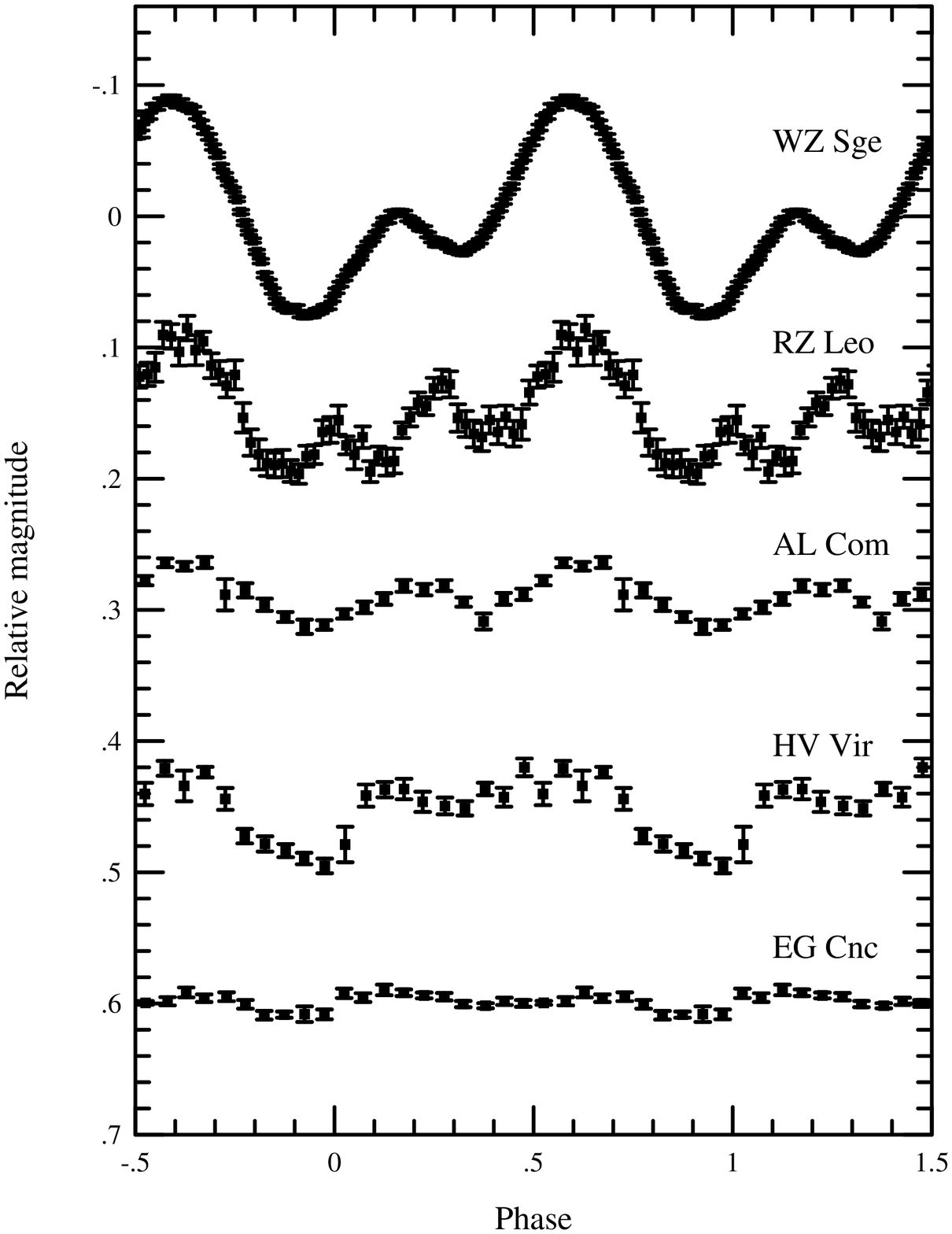}
  \end{center}
  \caption{Comparison of profiles of early superhumps of WZ Sge-type stars.
  The data are taken from: WZ Sge \citep{ish02wzsgeletter}; RZ Leo
  \citep{ish01rzleo}; AL Com \citep{kat96alcom}; HV Vir \citep{kat01hvvir};
  EG Cnc \citep{mat98egcnc}.  The phase in WZ Sge corresponds to the orbital
  phase.  Since the exact binary phases are unknown for the rest of the
  objects, the phases were arbitrarily taken so that the main hump maxima
  best match that of WZ Sge.  All stars exhibit characteristic double
  (or sometimes triple) humps with unequal amplitudes.
  }
  \label{fig:ecomp}
\end{figure}

   The most remarkable signature of WZ Sge-type outbursts is the
presence of ``early superhumps"\footnote{
  This feature is also referred to as {\it orbital superhumps}
  \citep{kat96alcom} or {\it outburst orbital hump} \citep{pat98egcnc}.
} during their earliest stage of superoutbursts.  Early superhumps have
a period extremely close to that of the binary period\footnote{
  The best-established case is the 2001 superoutburst of WZ Sge
  (\cite{ish02wzsgeletter} and references therein).  In a few other
  systems (e.g. AL Com: \cite{kat96alcom}, \cite{pat96alcom},
  \cite{nog97alcom}; EG Cnc: \cite{kat97egcnc}; \cite{mat98egcnc};
  \cite{pat98egcnc}), the periods of early
  superhumps have been found to be in good agreement with their quiescent
  photometric periods (most likely representing orbital periods).
}, and commonly show double-humped profile (figure \ref{fig:ecomp}; see
also \citet{kat96alcom} and \citet{kat98super} for detailed discussions),
in contrast to ordinary superhumps of SU UMa-type dwarf novae
[see \cite{war95suuma} for basic properties of SU UMa-type dwarf novae].
Early superhumps are the most discriminative feature of WZ Sge-type outbursts,
and have not been detected in other dwarf novae (see
\cite{kat01wxcet} and \cite{kat01hvvir}).
Although several models have been historically
proposed [e.g. enhanced hot spot \citep{pat81wzsge}, immature form of
superhumps \citep{kat96alcom}, jet or thickened disk
\citep{nog97alcom}], none of them has successfully explained all the
observed features of early superhumps.

\section{Recent Observations of Early Superhumps}
   During the early stage of the 2001 outburst of WZ Sge (e.g.
\cite{ish02wzsgeletter}), several authors (\cite{ste01wzsgeiauc7675};
\cite{bab01wzsgeiauc7678}) detected two-armed arch-like structures on
Doppler tomograms\citep{DopplerTomography} of He\textsc{II} and
C\textsc{III}/N\textsc{III} emission lines.  The feature resembled
those of `spiral structure' seen on Doppler tomograms of the dwarf nova
IP Peg (\cite{ste97ippeg}; \cite{ste98ippegerratum}) in outburst.
The spiral structure in Doppler tomograms of IP Peg has been widely
believed to represent tidally induced spiral shocks or spiral waves
\citep{ste97ippeg}.  However, it is not clear whether the same
interpretation can apply to a completely different (very low mass-ratio,
$q=M_2/M_1 < 0.1$ in contrast to $q\sim 0.5$ in IP Peg, and very low-mass
transfer rate, several orders of magnitudes lower than that of IP Peg)
binary parameters of WZ Sge.

   At the epoch of the detection of two-armed arch-like structures in
WZ Sge, the star showed double-wave early superhumps \citep{ish02wzsgeletter}.
Although there have been suggestions that early superhumps in photometry and
two-armed structures on Doppler tomograms have the same origin
(\cite{ish02wzsgeletter}; \cite{bab02wzsgeletter}; \cite{kuu02wzsge}),
no promising idea has been proposed to explain these features in the same
scheme.

\section{Tidally Distorted Accretion Disks in Close Binaries}
   Most recently, an alternative idea has been proposed to explain
arch-like structures in the outbursting IP Peg disk, by considering the
irradiation of the elevated disk formed by the horizontal
convergence ($-div(\mbox{\bfseries\itshape{v}})$) of three-body orbits
\citep{sma01tidal}.  This interpretation
successfully reproduced the arch-like structures, and the observed strong
intensity in the region of (${\it v_x}>0, {\it v_y}>0$), which is difficult
to explain within the scheme of conventional tidally induced spiral shocks.
\citet{ogi01tidal} studied tidal distortion of fluid disks, and succeeded
in reproducing the basically same features.  \citet{ogi01tidal} showed,
in realistic fluid disks, that $m=2$ inner vertical resonance play
a more important role even in high $q$ binaries.

\begin{figure*}
  \begin{center}
    \FigureFile(60mm,60mm){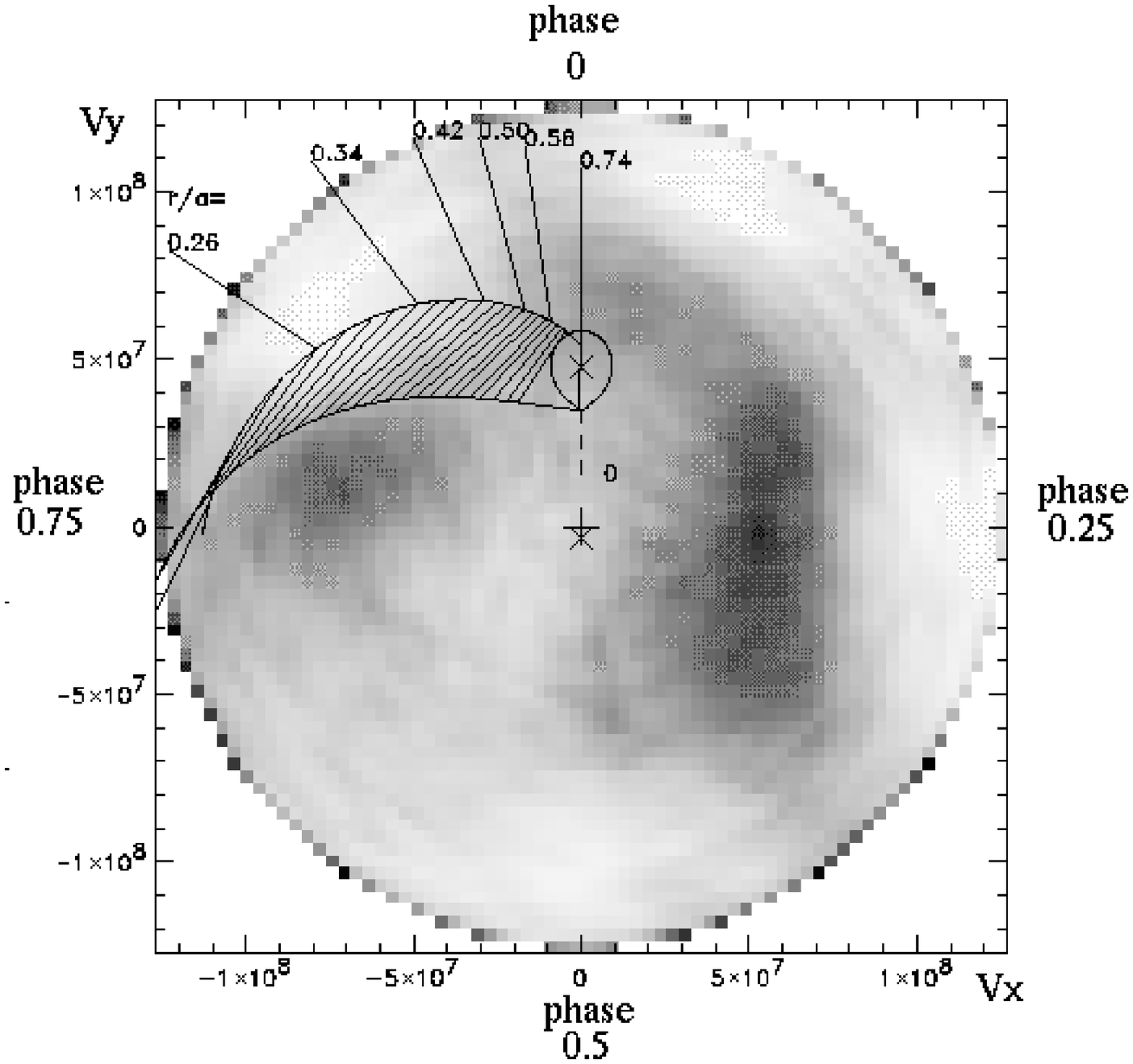}
    \FigureFile(50mm,60mm){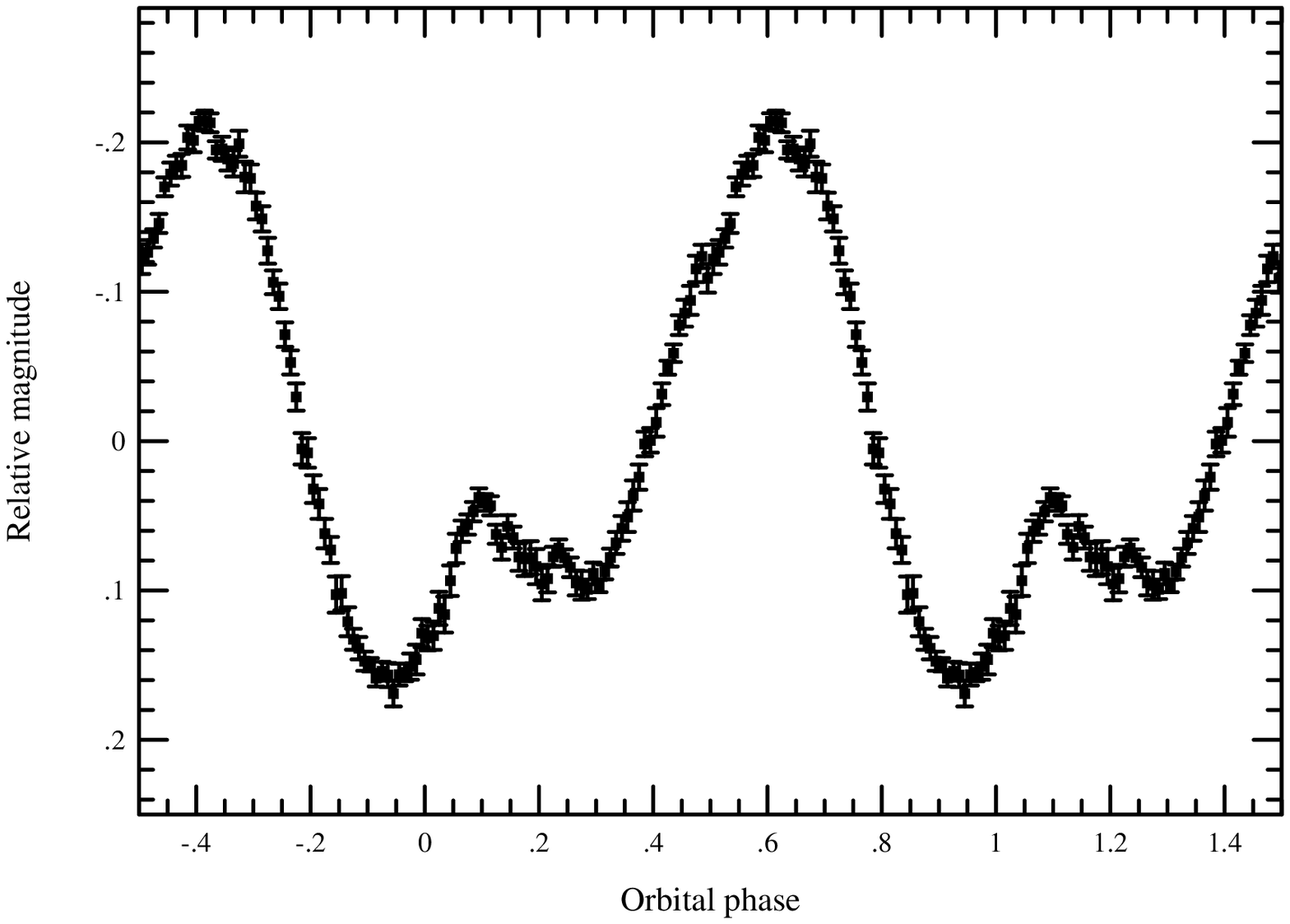} \\
   \vspace{0.1cm}
    \FigureFile(60mm,100mm){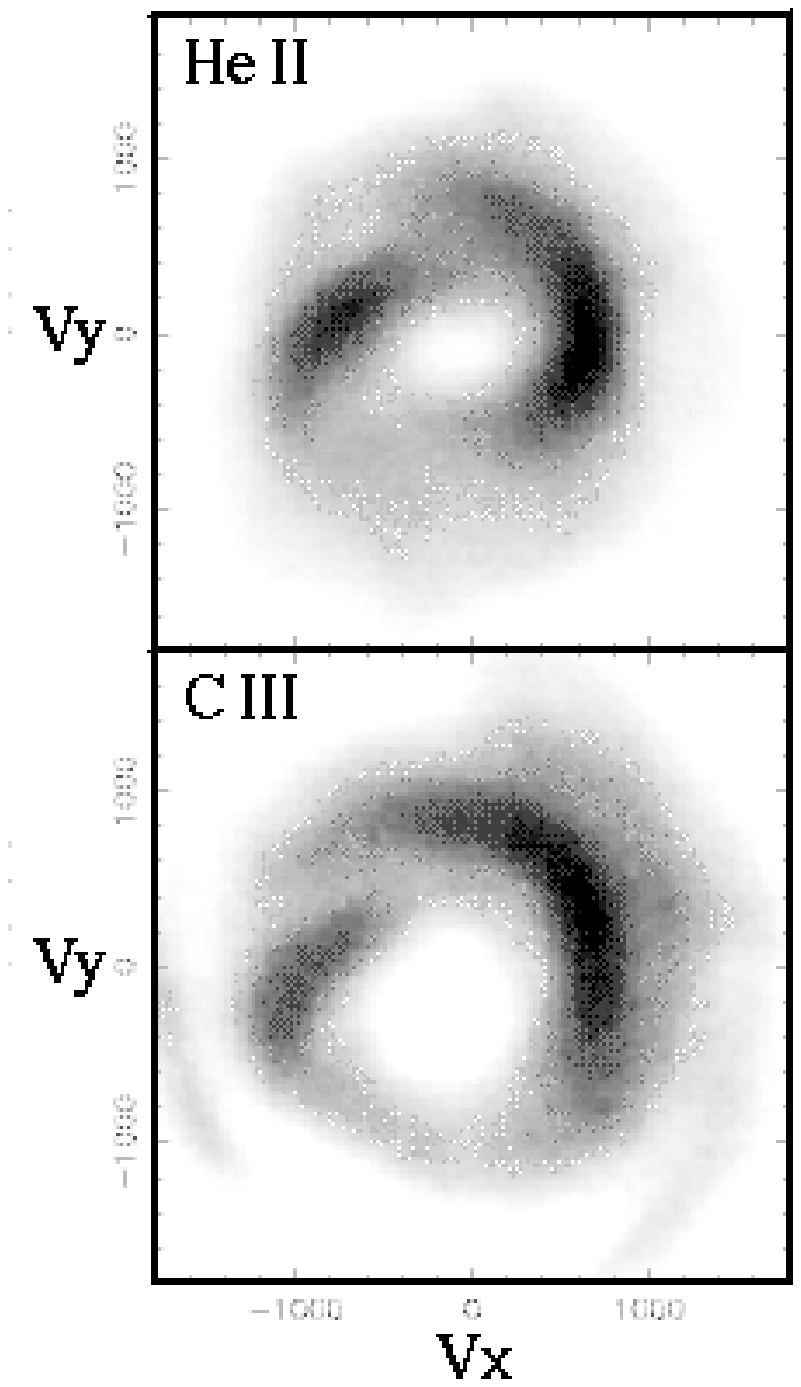}
    \FigureFile(50mm,60mm){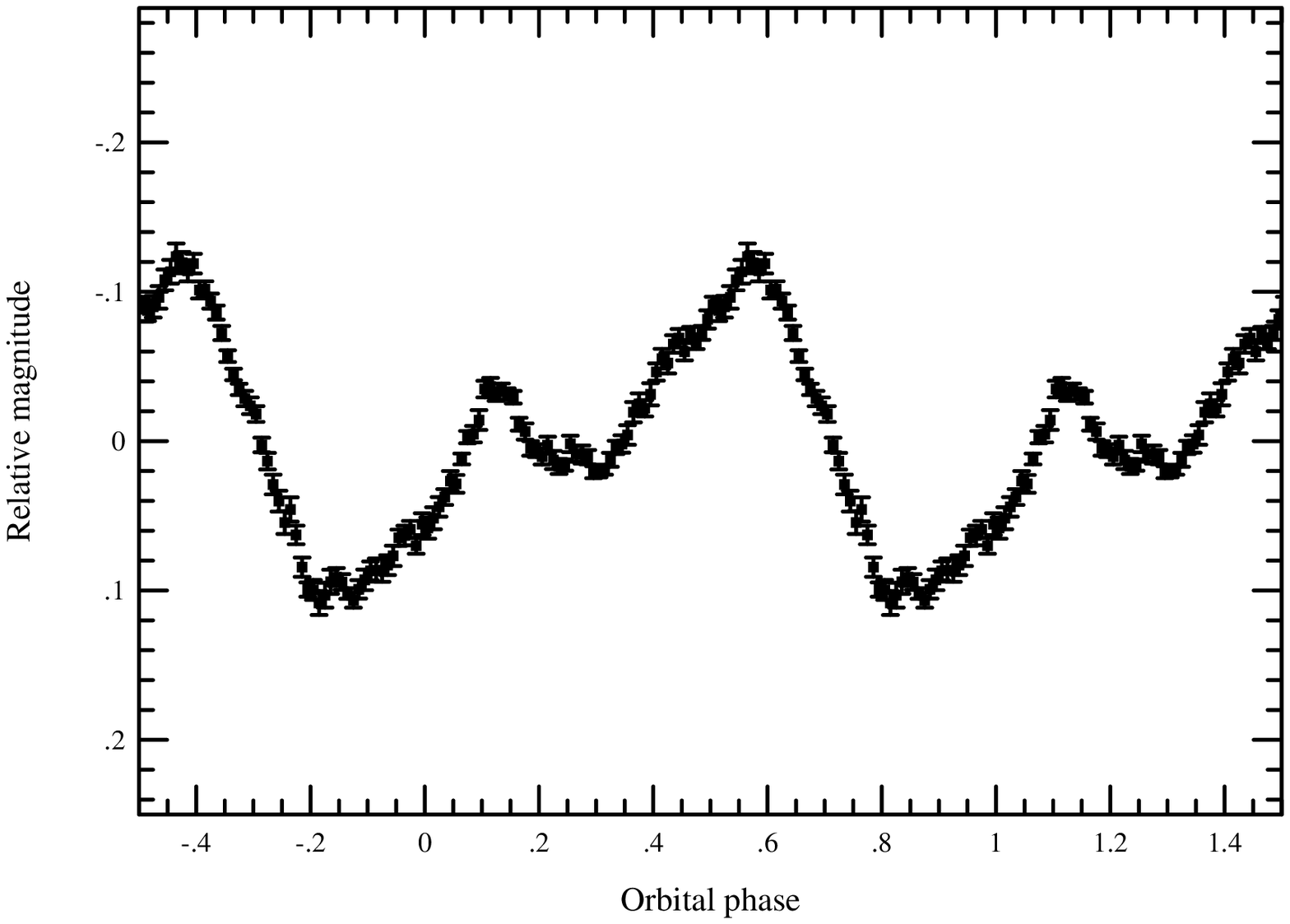}
  \end{center}
    \caption{(Upper left) Doppler maps of He\textsc{II} from
    \citet{bab02wzsgeletter} on 2001 July 24.  The Roche lobe and streamline
    were drawn with binary parameters of $\it{i}$=75$^{\circ}$ and $q$=0.071.
    The `phase' values represent the direction of the observer at given
    orbital phases.  The stronger He\textsc{II} feature is seen in
    the region with ${\it v_x}>0$, i.e. facing the observer near the
    binary phase 0.1--0.4.
    (Upper right) Phase-averaged light curve of early
    superhumps on the same night (the data are from \citet{ish02wzsgeletter}).
    The light curve shows double maxima of different amplitudes.
    The stronger hump is seen near the binary phase 0.6--0.7, almost
    opposite to what is expected from the Doppler tomogram.
    (Lower left) Doppler maps of He\textsc{II} and C\textsc{III} from
    \citet{kuu02wzsge} (the data were taken by D. Steeghs on 2001 July 28).
    Two-armed arch-like structures are clearly seen.  The asymmetry of
    the intensities of between arches is stronger in C\textsc{III}.
    (Lower right) Phase-averaged light curve on July 28.  The main maximum
    moved to phase 0.5--0.6, which is in agreement with the observed
    counterclockwise rotation of the features on Doppler maps since July 24.
    }
  \label{fig:dopcompare}
\end{figure*}

\section{New Interpretation of Early Superhumps}
   In low $q$ systems like WZ Sge, tidal force of the secondary is so weak
on a compact disk that a condition is usually achieved that angular
momentum is more effectively transported by viscosity than tidal force
of the secondary \citep{lin79lowqdisk}.  However, during the early stage
of vigorous outbursts of WZ Sge-type stars, the disk can sufficiently
expand \citep{osa95wzsge}.  In usual SU UMa-type dwarf novae, the growth
of 3:1 resonance, which is responsible of superhumps (\cite{whi88tidal};
\cite{hir90SHexcess}; \cite{lub91SHa}), is rapid enough to effectively
truncate the disk at the resonance radius.  In low $q$ WZ Sge-type stars,
the slow growth [the growth rate being proportional to $q^2$
\citep{lub91SHa}] can allow the disk to expand beyond the 3:1 resonance.
\footnote{
   Some observational implications of WZ Sge-type stars have been proposed
   to be a consequence of the disk beyond the 3:1 resonance
   \citep{kat98super}.
}
In such a condition, tidal distortion of the outer edge of the accretion
disk, as proposed by \citet{sma01tidal} and \citet{ogi01tidal}, can
effectively work, as shown in figure 3, $q$=0.1 case of \citet{ogi01tidal}.
This effect is predominant in the outer edge of the
accretion disk, which can explain stronger and distorted emission lines
in low velocity regions (e.g. \cite{bab02wzsgeletter}).  Since high-excitation
emission lines like He\textsc{II} require a high temperature ($\sim$ 25000 K),
the concentration of emission lines and arch-like patterns in low velocity
regions is otherwise difficult to explain \citep{sma01tidal}.  The arch-like
pattern in Doppler tomograms in WZ Sge is also stronger in the region of
(${\it v_x}>0, {\it v_y}>0$), which also can be naturally explained
as a result of stronger tides near the secondary, as in \citet{sma01tidal}
and \citet{ogi01tidal}.

   The maximum phases of early superhumps have been also problematic.
Early superhumps in WZ Sge-type dwarf novae consist of two maxima of
different amplitudes (figure \ref{fig:ecomp}).  In WZ Sge, the brighter
maximum is around binary phase 0.6--0.7 \citep{ish02wzsgeletter}.
While models involving an enhanced hot spot as in \citet{pat81wzsge}
would require such a maximum corresponding to the orbital phase of
a stronger emission feature, the available observation shows a complete
phase reversal (figure \ref{fig:dopcompare}).
Such a reversal can be naturally understood if one considers the source
of emission lines and stronger continuum is an irradiated
surface (\cite{sma01tidal}; \cite{ogi01tidal}) of the accretion disk.
[Since CVs are known to emit most of their energy in the UV wavelengths
(e.g. \cite{war95book}), re-emission of a small fraction of intercepted
UV photons in the optical wavelengths is expected to explain the observed
amplitudes of continuum variation (early superhumps)].
The decrease in amplitude of early superhumps in accordance with the
fading of the outburst (e.g. \cite{ish02wzsgeletter}) can be also
naturally understood as an effect of shrinkage of the disk, resulting in
decreasing tidal effects.

   \citet{ish02wzsgeletter} also showed that the exact period of early
superhumps is very slightly (0.05\%) shorter than the orbital period.
This effect was observed as a phase-shift of maxima of early superhumps.
This change looks likely to correspond to a counterclockwise rotation of
the patterns on Doppler tomograms at different epochs
(figure \ref{fig:dopcompare}).  The rotation of patterns may reflect
the variation of the direction of the `maximum height distortion' toward
a smaller disk radius (see figure 2 of \cite{ogi01tidal}).  Further
detailed calculations of a time-evolution of the outbursting WZ Sge disk
would provide a stringest test to this interpretation.

   Most recently, \citet{osa01wzsgehump} explained early superhumps
(which they call {\it early humps}) by considering the manifestation of
the tidal 2:1 resonance in accretion disks of binary systems with extremely
low mass ratios.  The present interpretation is different from that
of \citet{osa01wzsgehump} in its basic mechanism: \citet{osa01wzsgehump}
requires tidal {\it dissipation} while the present interpretation only
requires tidal {\it distortion} of a steady flow (hence it does not require
a shock).  The present interpretation has an advantage over all existing
models in its simplicity of the assumed physics and in that it can also
explain two-armed distortion patterns on Doppler tomograms as well as
photometric signals.

   The reason why such features are only observed in WZ Sge-type stars
can be explained as a combination of the large stored mass in quiescence
\citep{osa95wzsge} and the delayed exertion of the 3:1 resonance, which
is only reasonably achieved in extremely low $q$ and low mass-transfer
rate (and presumably low quiescent viscosity) systems as in WZ Sge-type
stars.  Since three-dimensional fluid calculation by \citet{ogi01tidal}
showed that $m=2$ inner vertical resonance can effectively work in relatively
high $q$ systems, we can expect similar phenomena expected in
\citet{lin79lowqdisk} in larger $q$ systems, which cannot hold particle
orbits of the 2:1 resonance within the Roche lobe.  Although
\citet{osa01wzsgehump} also presented an idea of distinguishing
WZ Sge-type stars from other CVs based on resonances involved, the present
interpretation is different from that of \citet{osa01wzsgehump} in that
\citet{osa01wzsgehump} only considered resonances in {\it Paczynski type}
orbits (e.g. \cite{ogi01tidal}), while the present interpretation involves
the three-dimensional effect as presented by \citet{ogi01tidal}.
The appearance of early superhumps and the reproduction of WZ Sge-type
outburst in RZ Leo \citep{ish01rzleo}, a binary with $q=0.14$, which is
approximately twice larger than those of other known WZ Sge-type stars,
can be understood as one of the most distinct consequences between the
present interpretation and that of \citet{osa01wzsgehump}.

\vskip 3mm

   The author is grateful to H. Baba, R. Ishioka and D. Steeghs for
allowing the use of their figures.  The author is also grateful to
Y. Osaki for drawing attention to the paper by \citet{lin79lowqdisk}.
This work is partly supported by a grant-in aid (13640239) from the
Japanese Ministry of Education, Culture, Sports, Science and Technology.

\end{document}